\documentclass[twocolumn,showpacs,aps]{revtex4}
\usepackage{graphicx}
\usepackage{pstricks}
\usepackage{dcolumn}
\usepackage{bm}
\usepackage{epsf}
\usepackage{epsfig}
\usepackage{amsmath,amsfonts,amsthm, amssymb,latexsym}
\usepackage{enumerate}

\newcommand{\beq}{\begin{equation}}
\newcommand{\eeq}{\end{equation}}
\newcommand{\bcn}{\begin{center}}
\newcommand{\ecn}{\end{center}}

\newcommand{\lsim}{\lower0.5ex\hbox{$\; \buildrel < \over \sim \;$}}
\usepackage{multirow}

\begin{document}

\title{Thermal Relaxation and Cooling of Quark Stars with a Strangelet Crust} 

\author{Joas Zapata}
 \email{jonathanzapata@id.uff.br}
\affiliation{Universidade Federal Fluminense--UFF, Niterói-RJ, Brazil}

\author{Rodrigo Negreiros}
 \email{rnegreiros@id.uff.br}
\affiliation{Universidade Federal Fluminense--UFF, Niterói-RJ, Brazil}
        
\author{Thiago Sales}
 \email{thiagosales@id.uff.br}
\affiliation{Universidade Federal Fluminense--UFF, Niterói-RJ, Brazil}

\author{Prashanth Jaikumar}
 \email{prashanth.Jaikumar@csulb.edu}
\affiliation{Department of Physics and Astronomy, California State University Long Beach, Long Beach, CA 90840, USA}

\date{\today}

\begin{abstract}
In this article, we explore the cooling of isolated quark stars. These objects are structured of a homogeneous quark matter core and crusted by matter. To do this, we adopt two kinds of crust: (i) a crust made of purely nuclear matter following the Baym-Pethick-Sutherland (BPS) equation of state (EoS) and (ii) a crust made of nuggets of strange quark matter (strangelets). Both models have the same quark matter core described by the MIT bag model EoS. Our main purpose is to quantify the effects of a strangelet crust on the cooling and relaxation times of these strange stars.  We also perform a thorough study of the thermal relaxation of quark stars, in which we have found that objects with a strangelet crust have a significantly different thermal relaxation time.  Our study also includes the possible effects of color superconductivity in the quark core.
\end{abstract}

\maketitle
\section{Introduction} \label{sec:Int}
According to the strange quark matter (SQM) hypothesis~\cite{bodmer1971, Terazawa1979, Witten1984}, strange matter which contains roughly equal numbers of up, down, and strange quarks may be the true ground state of the strongly interacting matter.
If the SQM hypothesis is true, many neutron stars could be in fact strange stars, i.e. large (dimensions $\sim$ km) compact stars made entirely of strange quark matter~\cite{Baym1976,alcock1986strange,weber2005strange}. In this work we assume the SQM hypothesis and consider compact stars that are made up of absolutely stable strange matter.

Given the self-bound nature of SQM~\citep[][]{Alcock1988}, many authors (see for instance~\cite{weber2005strange} and references therein), have considered the possibility of strangelets (droplets, or nuggets of SQM). Particularly, the authors of~\citep[][]{Farhi1984,BergerFarhi1987} assumed that strangelets are uniformly charged (i.e., constant chemical potential within the quark matter strangelet) to calculate a mass formula from them. This approach, however, has later been proved to be inconsistent as the electrostatic potential increases towards the strangelet's center thus causing quarks to migrate due to the resulting electric field. Improving on the work of~\citep[][]{Farhi1984,BergerFarhi1987} Heiselberg~\cite{Heiselberg1993} took into account the screening effect in strangelets and found a more accurate mass formula. He showed that the charge density is found to vary on a scale of the order of the Debye screening length $\lambda_{D}\sim 5~fm$ for strangelets with mass number $A\lesssim 10^{5}$.
As shown by Alford~\cite{Alford2008} the Debye screening plays a major role in the internal energy of strangelets, as it shuffles the electric charge.

If the SQM hypothesis is true, strange stars would be a new class of astronomical compact objects. Several physical scenarios have been proposed and theoretically discussed as to such possibilities~\cite{Witten1984, Baym1976,weber2005strange,Haensel1986,Alcock1986}. Relevant to the work we present here is the research of Alcock et al.~\cite{Alcock1986}, in which they considered the possibility for a strange star to maintain a thin crust of normal matter. They pointed out that the crust was mainly influenced by two factors: (\textit{i}) the tunnel effect through which ions might penetrate the core-crust gap and, (\textit{ii}) that the density at the base of the crust can not be denser than the neutron drip ($\epsilon_{drip}$) since free neutrons would come out of nuclei and fall into the strange core~\cite{Alcock1986,glendenning1992astrophys,Glendenning1995,Huang1997}. The latter consideration was revised by Huang and Lu~\cite{Huang1997} where they found that the maximum density at the base of the crust is about $\sim \epsilon_{drip}/5$ giving a maximum mass of $\sim 3.4\times 10^{-6}M_{\odot}$ for the crust, which is about one order of magnitude smaller than what had been found before. In the traditional picture, the surface of a bare strange star has a sharp edge of thickness $\sim 1~fm$~\cite{Alcock1986}. Below the surface lies quark matter which on the outermost layer should be positively charged (due to exhaustion of massive strange quarks), and above which resides a cloud of electrons (that guarantees the star's charge neutrality)~\cite{Alcock1986,Stejner2005,Usov1997}.

It has been shown, however, that if the surface tension $\sigma$ of the interface between quark matter and the vacuum is less than a critical value $\sigma_{crit}$ then large lumps of strange matter become unstable against fission into smaller pieces~\cite{Jaikumar2006,Alford2006}. As a result, the lower density surface region is replaced by a ``mixed-phase'' involving nuggets (strangelets) of positively charged strange matter in a neutralizing background of electrons. Jaikumar, Reddy, and Steiner~\cite{Jaikumar2006}, assuming zero surface tension and neglecting Debye screening, estimated that the ``mixed-phase'' crust might be $40-100~m$ thick. Later, Alford and Eby~\cite{Alford2008} found that if the surface tension of quark matter is low enough, the surface of a strange star will be a crust consisting of a crystal of charged strangelets in a neutralizing background of electrons. They calculated the thickness of the crust taking into account the effects of surface tension and Debye screening of electric charges. Their results showed that the strangelet crust's size can range from zero to hundreds of meters thick and, the thickness is greater when the strange quark is heavier and the surface tension is smaller~\cite{Alford2008}. In this work we will further explore the possibility of a strangelet crust on strange stars and their implications to the thermal evolution of such stars.

Since the proposal of strange stars, many efforts have been devoted to indicate  observational properties (if any) that may be useful to distinguish strange stars from neutron stars, as they share many similar (observable) macroscopic properties (such as gravitational mass, for instance). One possibility to reach that goal is by their thermal evolution, as quark stars may exhibit a fairly distinct cooling as opposed to ordinary neutron stars. The cooling of neutron stars is dominated, mainly, by neutrino emissions for the initial $\sim 1000$ years, later being replaced by surface photon emissions~\cite{Page2006,Tsuruta1998}. Due to very different compositions/morphology between the neutron star core and crust, it takes $\sim 1-100$ years for the star to thermalize~\cite{Lattimer1994,Sales2020}.

The situation for crusted strange stars is significantly different since the presence of deconfined quark matter plays an important role in the cooling of the star~\cite{Blaschke2000, Grigorian2005, Blaschke2006}. In this article we will revisit the cooling of strange quark stars, considering the effects of a  strangelet crust as described by~\cite{Jaikumar2006,Alford2006}. We will compare our findings to the cooling of quark stars (QS) with nuclear matter crusts. Our main goal is to quantify the effects of a strangelet crust on the cooling calculation of quark stars. We will also study the thermal relaxation of quark stars, which to the extent of our knowledge has never been studied in details, therefore we study such properties here.

The remainder of this paper is organized as follows: in section~\ref{Model} we will describe the microscopic model for crusted strange stars and we present the results for the macroscopic structure of our models of quark stars. In section~\ref{Ther}, we will explore the thermal evolution of these stars and discuss their principal characteristics. Furthermore, we also include superconductivity and compare them with observations. Finally, the conclusions and perspectives will be presented in Sect.~\ref{Conclusions}. 

\section{MICROSCOPIC MODEL} \label{Model}
\begin{figure}[!ht]
 \centering
 \vspace{1.0cm}
 \includegraphics[angle=0,width=9 cm]{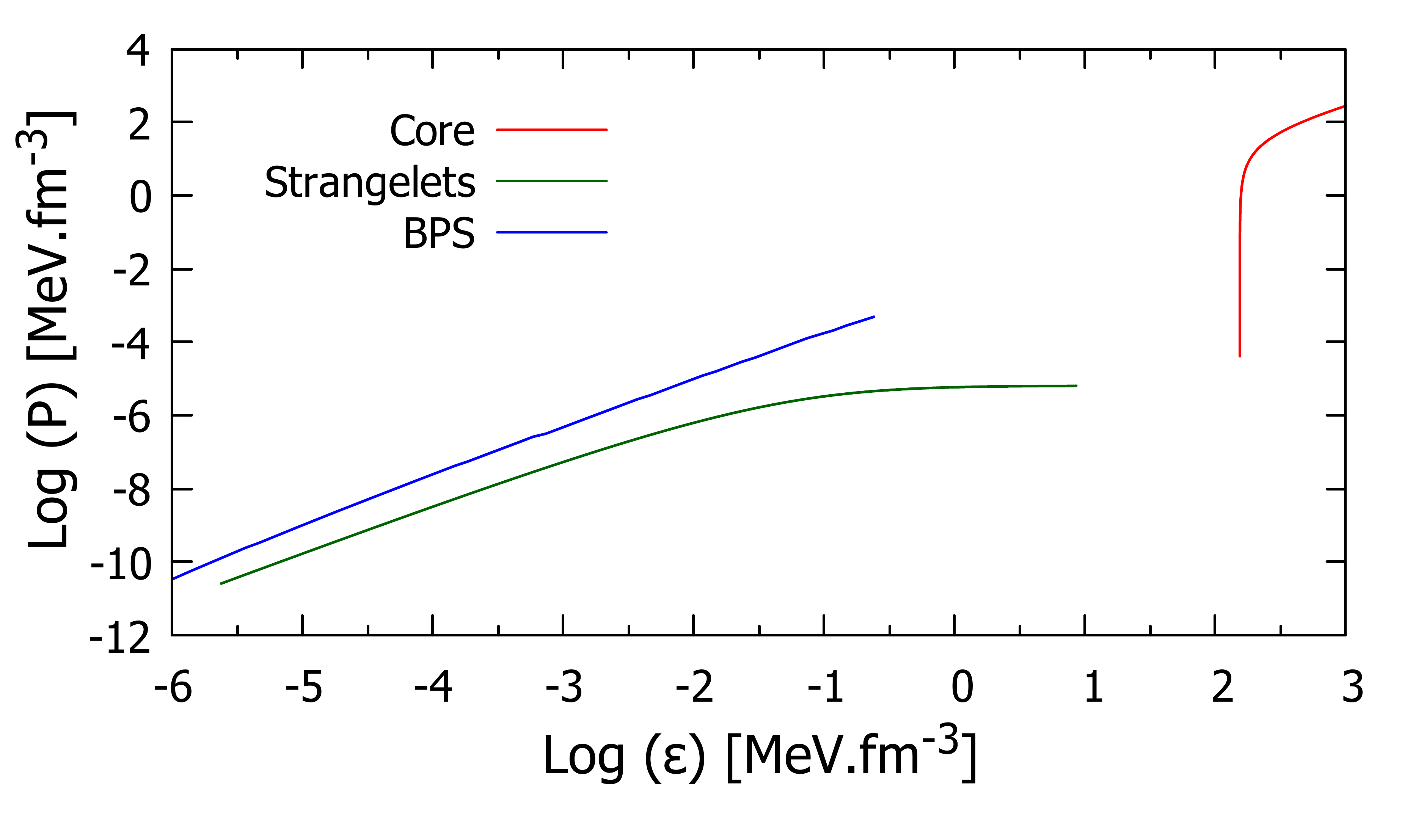}
 \vspace{-0.6cm}
 \caption{\label{fig:alleos}Equation of state for strange quark matter (Core) surrounded by (i) a nuclear BPS crust and (ii) a strangelet crust.}
 \end{figure}

The structure of a quark star studied here consists of two parts: the crust and the core of the star (obviously we are not considering bare quark stars, in which case they would not be crusted). The crust is characterized by a low density regime. Here, we consider two scenarios: (i) the (traditional) nuclear Baym-Pethick-Sutherland (BPS) equation of state~\cite{Baym1971, glendenning1992astrophys} - in which case the crust must necessarily have a maximum density limited by approximately the neutron drip density ($\epsilon_{drip}$); and, (ii) a strangelet crust as described by~\cite{Jaikumar2006}. That is, if the surface tension of the interface between quark matter and the vacuum is less than a critical value, then large lumps of strange matter become unstable against fission into smaller pieces; as a result the crust consists of a crystalline structure of charged spherical strangelets in a neutralizing background of electrons. For the star's core we adopt a traditional MIT bag model equation of state in which the parameters are set as $m_{s}= 100~MeV$, $B^{1/4} = 128.9~MeV$ and\textbf{ $\alpha_{s}=0.4$} (the strong interaction coupling constant). We note that such a model was chosen for its simplicity. It is important to mention that more sophisticated quark models have been proposed, such as the NJL model~\cite{Nambu1961,Nambu1961b}, PNJL model~\cite{Fukushima2003,Fukushima2004,Ratti2006} (and references therein) - these models, however, lead to a qualitatively similar composition, thus it is unlikely that they would strongly modify the thermal evolution, which is the focus of this research. The Figure~\ref{fig:alleos} shows the equation of state of a quark star with a nuclear (BPS) matter (labeled ``BPS'') and strangelet crust (labeled ``Strangelets''). The transition point between Core and crust occurs at \textbf{$\epsilon_{tr}\sim 153.76~MeV.fm^{-3}$}.

 \begin{figure}[!ht]
 \centering
 \vspace{1.0cm}
 \includegraphics[angle=0,width=9 cm]{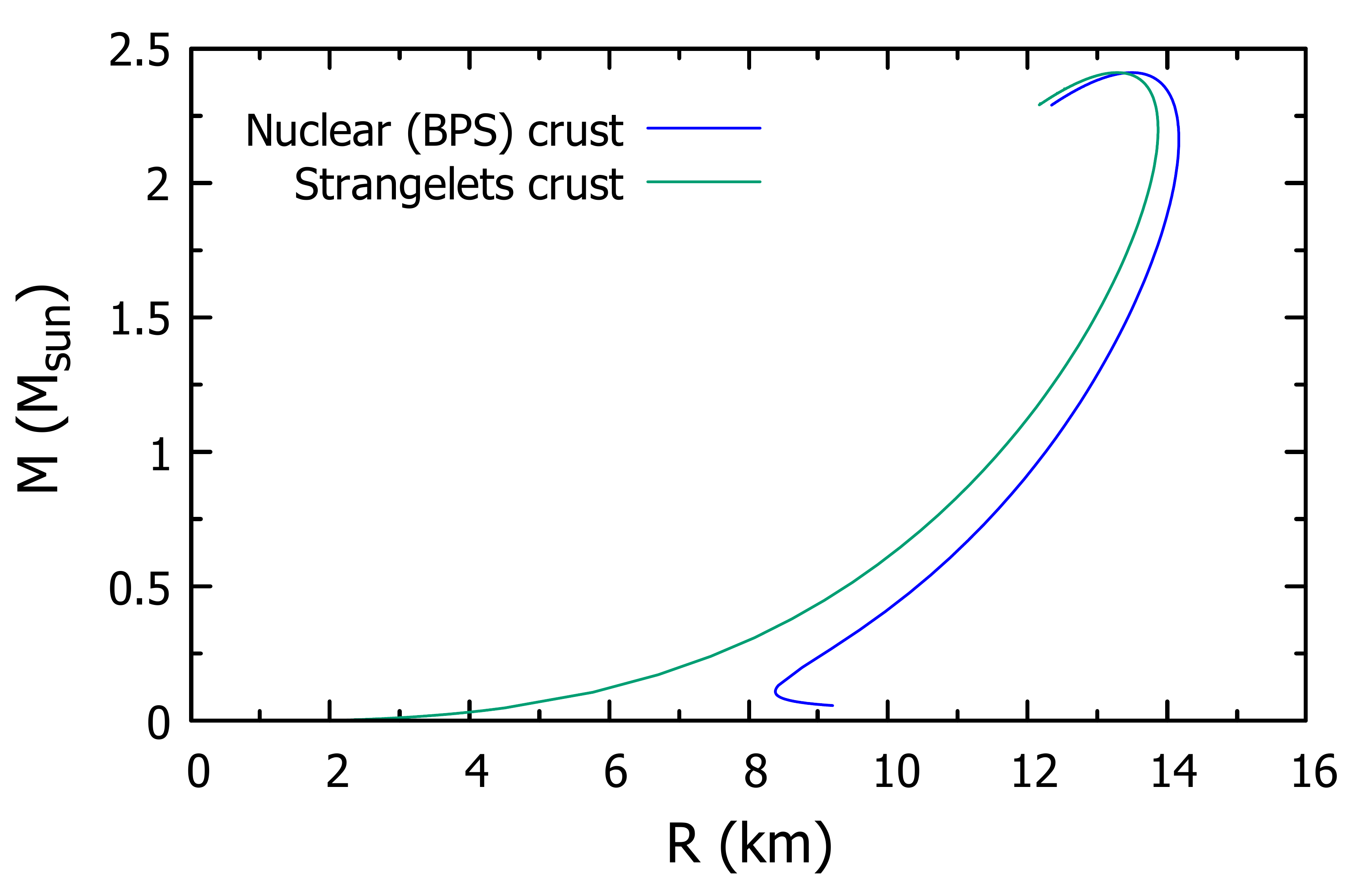}
 \vspace{-0.6cm}
 \caption{\label{fig:fam}Mass-radius diagram for quark stars whose equations of state are shown in Fig.~\ref{fig:alleos}. Both sequences have the same maximum mass $\sim 2.41M_{\odot}$}
 \end{figure}
\begin{table}[t!]
\caption{\label{table:stars} Properties of some quark stars from Fig.~\ref{fig:fam}. We differentiate them with the label strangelets or nuclear (BPS) crusts. $\epsilon_c$ is the central density.}
\begin{tabular}{cccc}
\hline\hline
\multirow{2}{*}{\begin{tabular}[c]{@{}c@{}}$\epsilon_{c}$\\ (MeV/fm$^{3}$)\end{tabular}} & \multirow{2}{*}{\begin{tabular}[c]{@{}c@{}}M\\ ($M_{\odot}$)\end{tabular}} & strangelet crusts & Nuclear (BPS) crusts \\ \cline{3-4} 
               &                & R (km)           & R(km)     \\ \hline
$237.24$       & $1.42$        & $12.78$          & $13.27$   \\ 
$257.49$       & $1.60$        & $13.19$          & $13.62$   \\ 
$288.59$       & $1.82$       & $13.55$          & $13.93$   \\ 
$327.36$       & $2.00$      & $13.77$          & $14.11$   \\ \hline\hline
\end{tabular}
\end{table} 
 
With the EoS's in Figure~\ref{fig:alleos}, we can solve the  Tolman--Oppenheimer-Volkoff (TOV) equations~\cite{Tolman1939, Oppenheimer1939} and find the structure of the quark stars. In Figure~\ref{fig:fam} we show the sequences of quark stars obtained from the EoS's. We note that, as expected, the only difference between the models studied is the description of the crust, both sequences have the same maximum mass, $\sim 2.41M_{\odot}$. Furthermore, due to the different crust modelling, we also see a significant difference between the stellar radius in each sequence. We summarize the macroscopic properties found for the two models studied in Table~\ref{table:stars}. The most notable distinction between them can be found as: quark stars with nuclear (BPS) matter crust have larger radii, crust thickness and mass $\Delta R_{crust} \sim 500~m$ and  $\sim 10^{-4}M_{\odot}$, respectively.
Quark stars with strangelet crust, on another hand, have smaller radii, crust thickness and mass $\Delta R_{crust}\sim 20~m$ and $\sim 10^{-5}M_{\odot}$, respectively. 

Later, in our study of these quark stars, we will consider the possibility of color superconductivity. The pattern that will be considered is the Color-Flavor-Locked (CFL) phase~\cite{Alford2001}, where all quarks of all colors are paired to form Cooper pairs. The CFL phase is the most likely condensation pattern at densities of $>2\epsilon_{0}$ (where $\epsilon_{0}$ is the nuclear  matter density)~\cite{Alford2008b}. Intermediate densities ($\sim 2\epsilon_{0}$) model calculations indicate that quark matter is in a 2SC phase~\cite{Alford2008b}. Another possibility is that quark matter forms a crystalline superconductor, where the momenta of the quark pairs do not add to zero~\cite{Alford2001b, Bowers2002}. Given the densities of the quark cores in our model, we will consider only the CFL phase. It should be noted that one expects corrections to the quark matter EoS when pairing is present, however, the effects of such corrections to the structure of the star are only noticeable for pairing gaps $\Delta \gtrsim 50$ MeV~\cite{Alford2003}. Therefore, for the values of $\Delta$ considered here ($0.1-10$~MeV) they can be safely ignored. Our study is still valid for any quark pairing scheme (not necessarily color superconductivity), as long as it affects all quark flavors in a similar way~\cite{Rodrigo2012}. In the next section, we will analyze the thermal evolution of our two models of quark stars from Table~\ref{table:stars} and thereby determine the differences between them. We also consider the superfluidity possibility and the thermal relaxation analysis. Our results will be compared with the prominent thermal observations. 
\section{Cooling}\label{Ther}
In this section, we study the thermal evolution of our two models discussed above. The cooling of a compact star is governed by the general relativistic thermal balance, and transport equations given by $(G = c =1)$~\cite{thorne1977, Riper1991, weber2017pulsars}
\begin{eqnarray}
\frac{\partial(le^{2\phi})}{\partial m}&=& - \frac{1}{\epsilon\sqrt{1-2m/r}}\left(\epsilon_{\nu}e^{2\phi}+c_{v}\frac{\partial(Te^{\phi})}{\partial t} \right),\label{thermal}\\
\frac{\partial(Te^{\phi})}{\partial m} &=& - \frac{(l e^{\phi})}{16\pi^{2}r^{4}\kappa\epsilon\sqrt{1-2m/r}},\label{transport}
\end{eqnarray}
where the macroscopic dependencies are: the radial distance $r$, the energy density $\epsilon(r)$ and, the stellar mass $m(r)$. Since the central star temperature at the beginning of its thermal life is not larger than $10^{11}$ K $\sim 1-10$ MeV, the effects of finite temperatures on the equation of state can be neglected to a very good approximation. Consequently,  TOV's equations do not depend on time and thus need to be solved only once - which is fortunate as the thermal and structural properties are then uncoupled. Moreover, the thermal properties are represented by the temperature $T(r,t)$, luminosity $l(r,t)$, neutrino emissivity $\epsilon(r,T)$, thermal conductivity $\kappa(r,T)$ and specific heat $c_{v}(r,T)$. The boundary conditions of the Eqs.~\eqref{thermal} and~\eqref{transport} are determined both by the luminosity at the stellar center and at the stellar surface. The luminosity vanishes at the stellar center since at this point the heat flux is zero. At the surface, the luminosity is defined by the relationship between the mantle temperature and the temperature outside of the star~\cite{Page2006, Blaschke2000}. The microscopic input in the Eqs.~\eqref{thermal} and~\eqref{transport} are the neutrino emissivities, specific heat, and thermal conductivity. For the quark core, we consider the processes involving quarks: the quark direct Urca (QDU), quark modified Urca (QMU), and quark bremsstrahlung processes (QBM). If the electron fraction vanishes entirely in quark matter ($Y_{e}=0$, in the limit in which $m_s \rightarrow 0$), both the quark direct and the quark modified Urca processes become unimportant, and the neutrino emission is then dominated by bremsstrahlung processes only. The emissivities of such processes were calculated in~\cite{Iwamoto1982}, we use the specific heat for the quark phase as calculated in~\cite{Iwamoto1982} and, the thermal conductivity comes from~\cite{haensel1991}
\begin{figure}[!ht]
 \centering
 \vspace{1.0cm}
 \includegraphics[angle=0,width=9 cm]{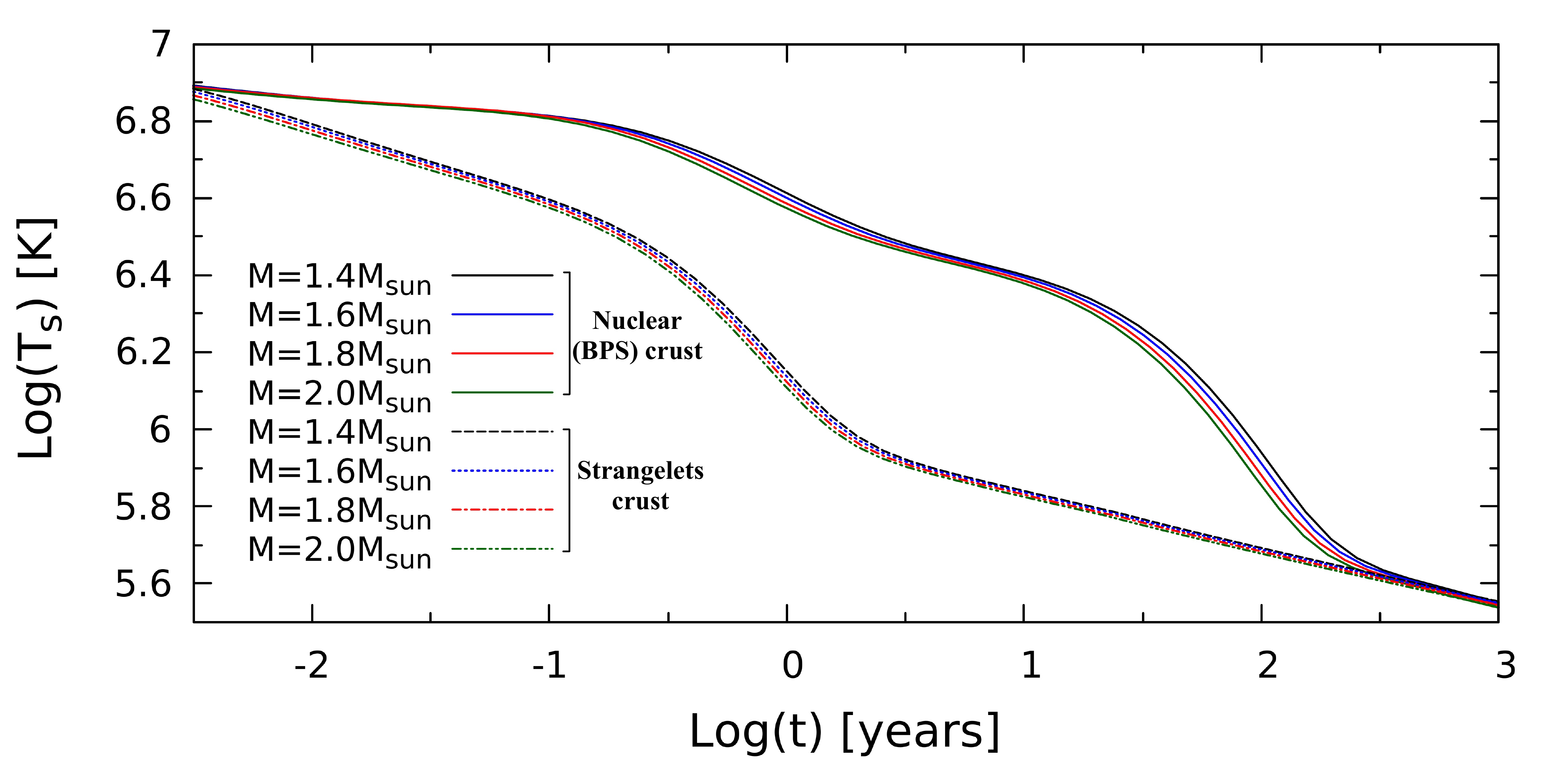}
 \vspace{-0.6cm}
 \caption{\label{fig:all} Cooling of quark stars with gravitational masses from Table~\ref{table:stars}. $T_{s}$ denotes the red-shifted temperature, and the x-axis the age, $t$, in years. Solid lines (upper band) represent quark stars with nuclear (BPS) crust and dashed lines (lower band) are stars with strangelet crust.}
 \end{figure}
 
To investigate the difference between a quark star with nuclear (BPS) crust and those with strangelet crust, we analyze the cooling of quark stars of the same mass for both. The thermal evolution of our models is illustrated in Figure~\ref{fig:all}, where we show a typical cooling curve, that is, the red-shifted surface temperature ($T_{s}$) as a function of the age ($t$) of the star. 
The results indicate that there is little difference between the cooling of stars with different gravitational masses within the same model, both for stars with nuclear (BPS) crust (solid lines) as well as for stars with strangelets (dashed lines) crust. Additionally, for each model, as the star's gravitational  mass increases, the surface temperature becomes slightly lower.  On the other hand, we can notice a significant difference when comparing the cooling behavior exhibited within each model. Most noticeably quark stars with strangelet crust cool down significantly faster than quark stars with nuclear (BPS) crust. We think that this is due to the thinner nature of the strangelet crust. 
\subsection{Thermal Relaxation}
\begin{figure}[!ht]
 \centering
 \vspace{1.0cm}
 \includegraphics[angle=0,width=9 cm]{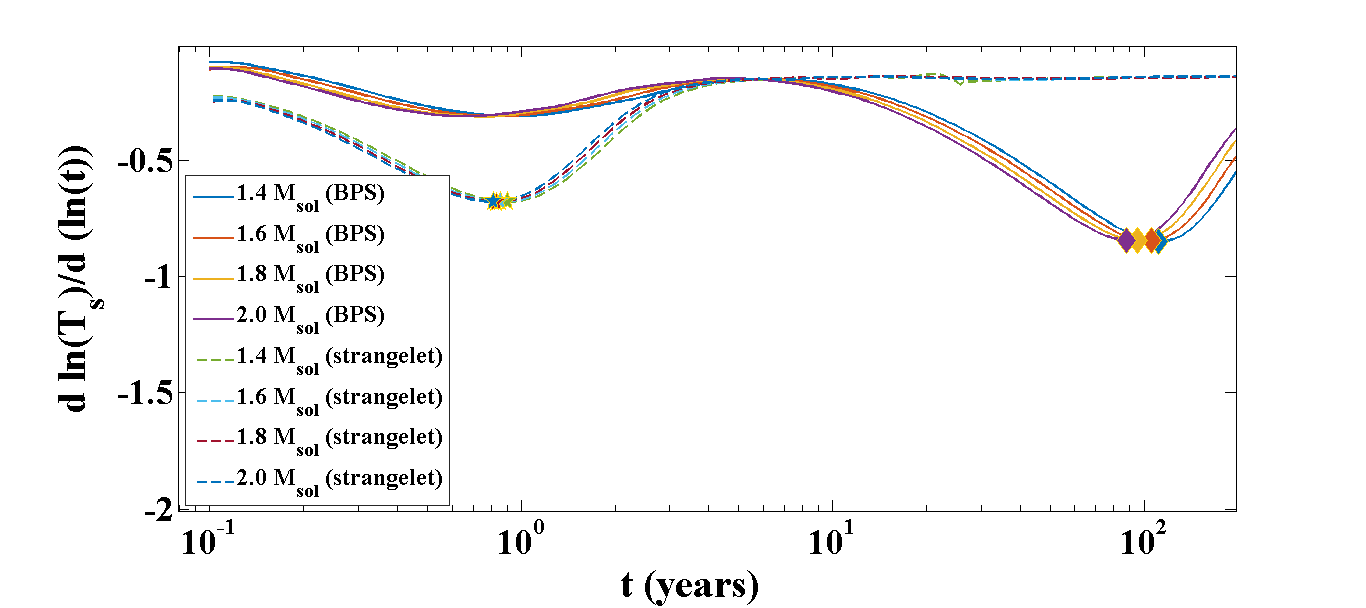}
 \vspace{-0.6cm}
 \caption{\label{fig:thermrelax} The $\ln{T_s}$ variation rate with respect to $\ln{t}$ versus ages for our quark stars from Fig.~\ref{fig:all}. Solid lines are quark stars BPS crusted and dashed lines are strangelet crusted. The highlighted diamond and star points represent the moment of the most negative slope, i.e, their relaxation times.}
 \end{figure}

\begin{figure}[!ht]
 \centering
 \vspace{1.0cm}
 \includegraphics[angle=0,width=9 cm]{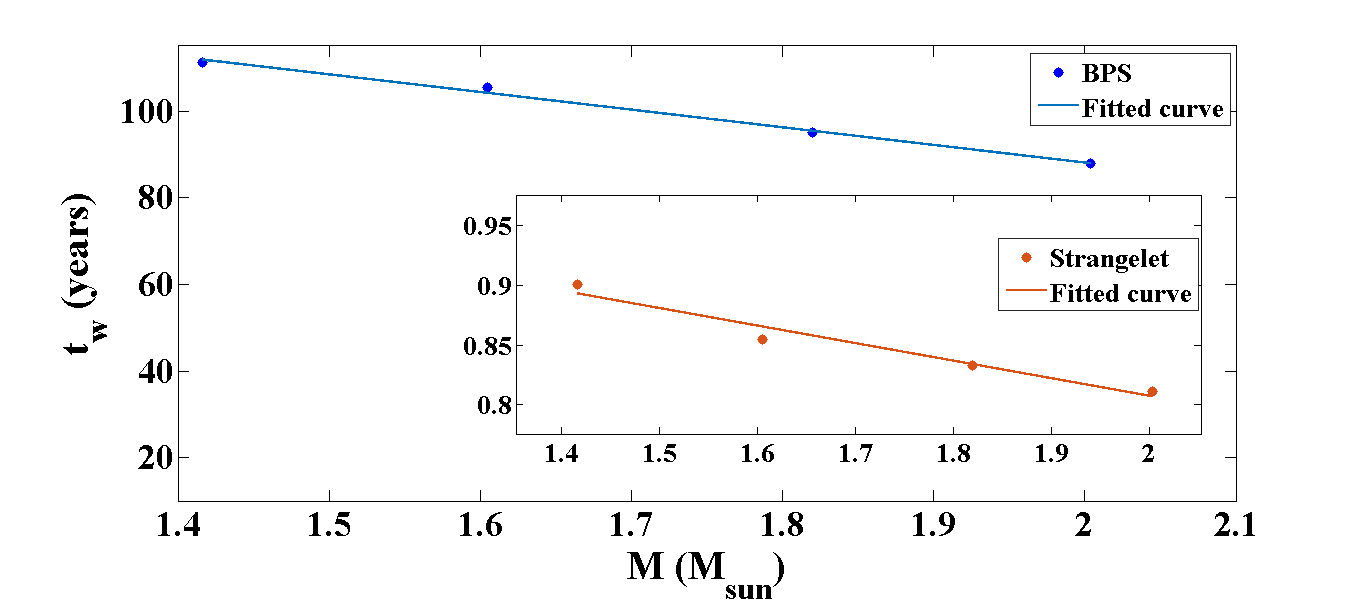}
 \vspace{-0.6cm}
 \caption{\label{fig:relax} Relaxation time vs gravitational mass. The blue line is for quark stars BPS crusted and the red line is for strangelet crusted. The two models have the same linear behavior, even though the gravitational mass increases while the relaxation time decreases.}
 \end{figure}

In order to quantify the faster cooling exhibited by quark stars with strangelet crusts we now discuss their thermal relaxation. As shown by Lattimer et al.~\cite{Lattimer1994} the thermal relaxation timescale $t_{w}$ is defined as the moment of the most negative slope of the cooling curve of a young neutron star. It is given in~\cite{Gnedin2001} by 
\begin{eqnarray}
t_{w}=max\left |\frac{dln(T_{s})}{d(ln(t))}\right |.
\end{eqnarray}
For ordinary neutron stars such relaxation times are typically between $\sim 10-100$ years. The  thermal relaxation time for ordinary neutron stars is determined mainly by the crust thickness $\Delta R_{crust}$, given in~\cite{Lattimer1994, Gnedin2001}; although it has been recently demonstrated that depending on how widespread (within the core of the star) the direct Urca process is, the thermal relaxtion time may be drastically larger~\cite{Sales2020}. We begin by showing in 
figure~\ref{fig:thermrelax} the $\ln({T_s})$ variation rate with respect to $\ln({t})$ for a representative sample of stars of the two models studied. Solid lines represent the quark stars with nuclear (BPS) crusts and dashed lines are those with strangelet crusts. Diamonds and stars indicate the maximum absolute value of each curve, thus representing the relaxation time. We now have a quantitative measure of how fast quark stars with strangelet crusts cool down with respect to those with BPS crusts. We perceive that the relaxation times of quark stars with strangelet crust ($\sim 1$ year) are two orders of magnitude smaller than quark stars with BPS crust. We can also obtain a direct relation between the relaxation time of quark stars and their masses (much like that obtained for ordinary stars~\cite{Sales2020}), which is shown in Fig.~\ref{fig:relax}. We observe that the relaxation time exhibits a linear behavior for both models, with the slope of the curve mainly dependent on the average thickness of the crust in each model.

\subsection{Superfluidity Effects}
As mentioned in~\cite{Rodrigo2012} and references therein, we expect strange matter to be in a superconducting phase. The most likely condensation pattern for strange matter in the high density cores of quark stars is that of {\it Color Flavor Locking} (CFL)~\cite{Alford2001}, in which all quarks are paired. Because of pairing, the direct Urca process is suppressed by $e^{-\Delta/T}$ factor, and the modified Urca and the Bremsstrahlung process by $e^{-2\Delta/T}$ factor, for $T\leq T_{c}$, where $\Delta$ is the gap parameter for the CFL phase and, $T_{c}$ is the pairing critical temperature~\cite{Alford2001,Alford2008b}. Moreover the specific heat of quark matter is also modified by the factor $3.2(T_{c}/T)[2.5-1.7(T/T_{c})+3.6(T/T_{c})^{2}]e^{-\Delta/T}$~\cite{Blaschke2000}. The critical temperature for the CFL phase is currently not known, however, it is believed to be smaller than the standard Bardeen-Cooper-Schrieffer ($T_{c}\simeq 0.57\Delta$), due to instanton-anti-instanton effects~\cite{Blaschke2000, Rodrigo2012}. Here, we use $T_{c}\simeq 0.4\Delta$.

\begin{figure}[!ht]
 \centering
 \vspace{1.0cm}
 \includegraphics[angle=0,width=9 cm]{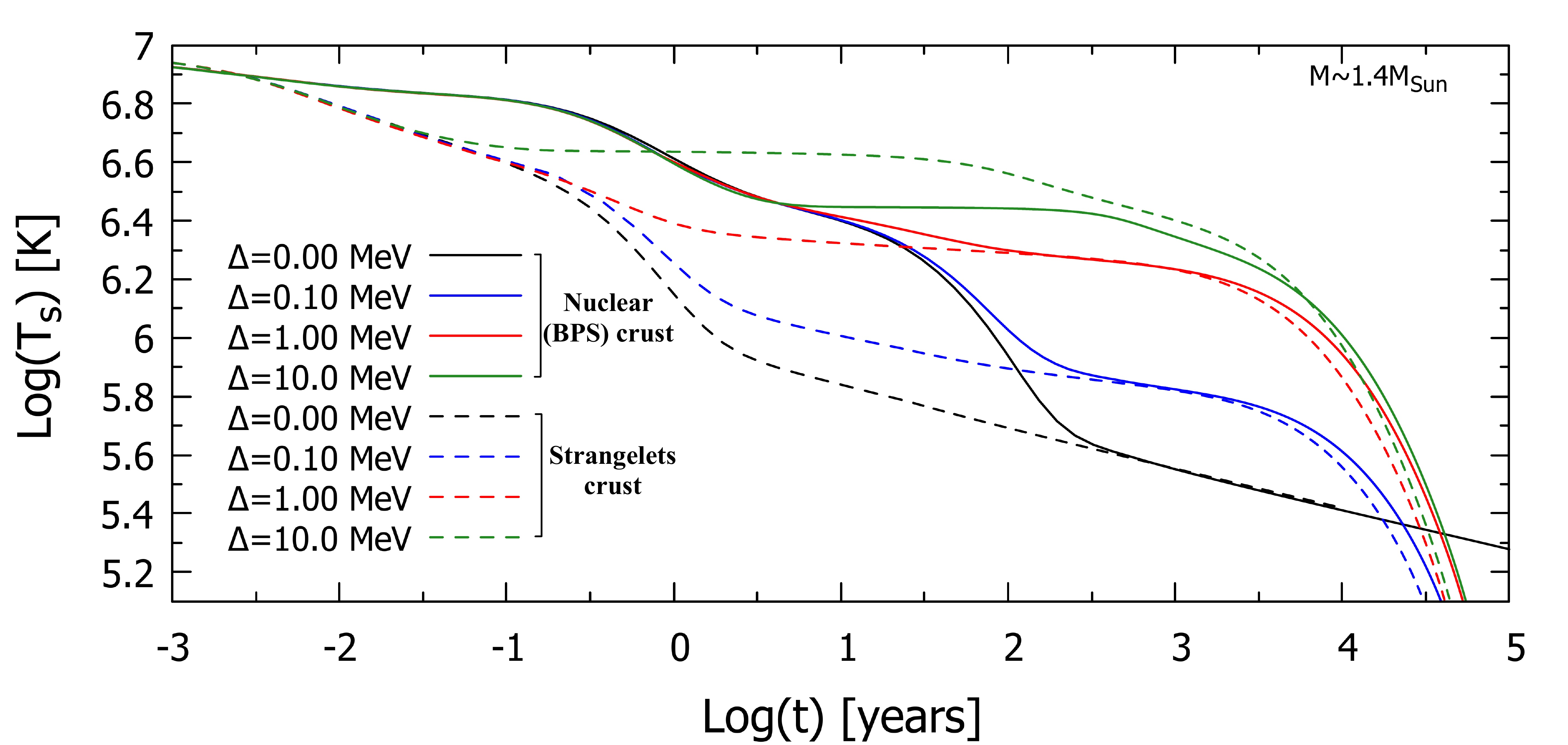}
 \vspace{-0.6cm}
 \caption{\label{fig:14} Cooling of quark stars with a gravitational mass of $\sim 1.4M_{\odot}$. $T_{s}$ denotes the temperature as measured by a distant observer, and the x-axis the age in years. Solid lines represent quark stars with nuclear (BPS) crusts, and dashed lines are stars with strangelet crusts, for different values of the CFL gap ($\Delta$).}
 \end{figure}
 \begin{figure}[!ht]
 \centering
 \vspace{1.0cm}
 \includegraphics[angle=0,width=9 cm]{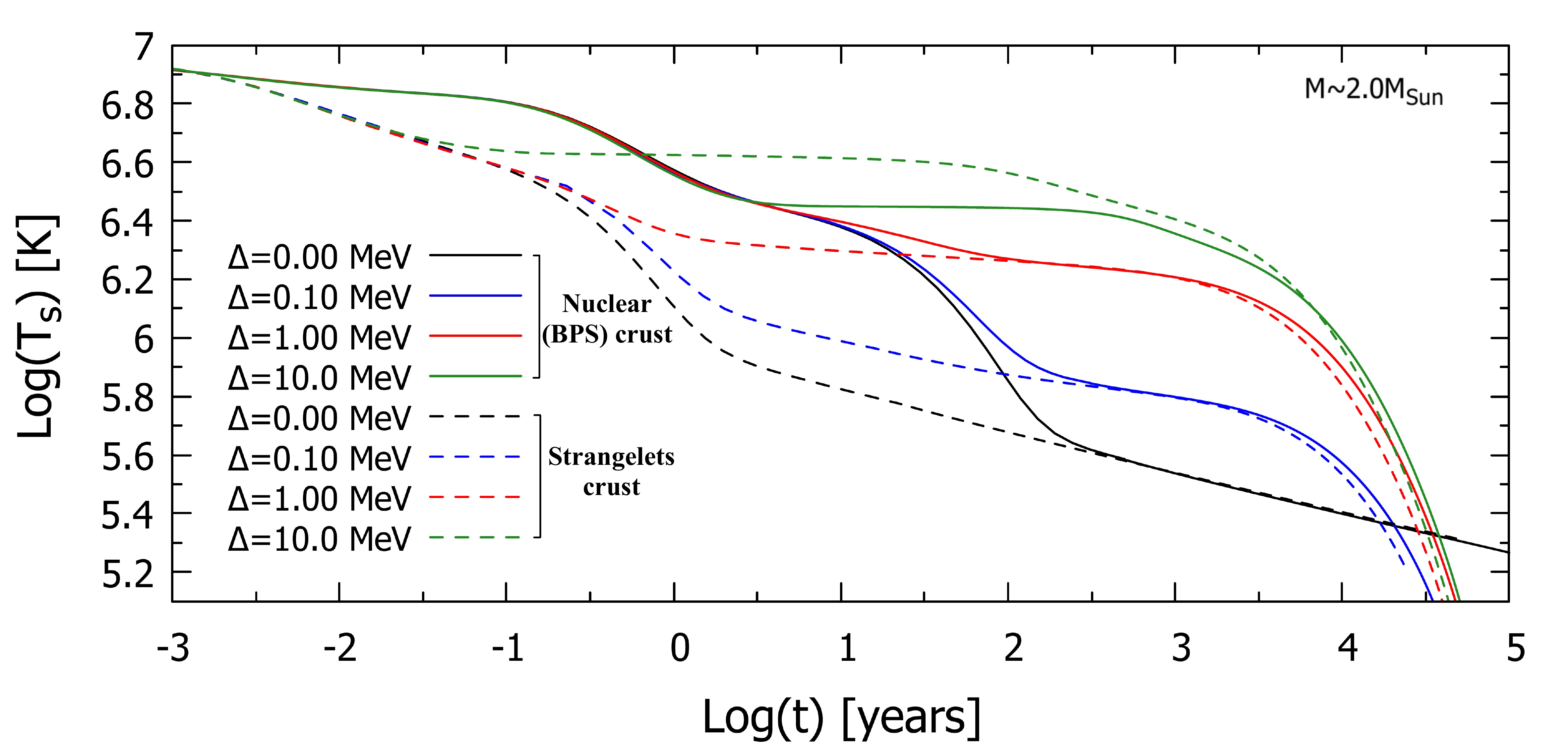}
 \vspace{-0.6cm}
 \caption{\label{fig:20} Cooling of quark stars with a gravitational mass of $\sim 2.0M_{\odot}$. $T_{s}$ denotes the temperature as measured by a distant observer, and the x-axis the age in years. Solid lines represent quark stars with nuclear (BPS) crusts, and dashed lines are stars with strangelet crusts, for different values of the CFL gap ($\Delta$).}
 \end{figure}
We have plotted the cooling curve for quark stars whose quark core is only composed of strange quark matter in the CFL phase, for different values of gap ($\Delta$) in the Figs.~\ref{fig:14}-\ref{fig:20} and, we compare them with stars without superfluidity. These quark stars have masses of $\sim 1.4M_{\odot}$ and $\sim 2.0M_{\odot}$, respectively. In this paper we limit our study to pairing with small gaps, given by $\Delta =0.1,~1.0,~10$ MeV.  We have not considered the cooling from processes involving the Goldstones bosons in the CFL phase. Although these processes are important for the core, they are not effective at cooling stars with a crust, and thermal relaxation of the crust is still the key factor.
We  note  a very distinctive behavior, depending on the value chosen for the superconductivity gap.
We see that objects  with a higher $\Delta$, thus stronger pairing, will result in slower cooling. 
For completeness we have also studied scenarios in which $\Delta\geq 10~MeV$ and have found that the resulting thermal evolution is essentially the same $\Delta=10~MeV$. This comes from the fact that the exponential $exp(-\Delta/T)$ effectively saturates for $\Delta>10~MeV$.  Here we note that superconductivity effects were only considered in the quarks at the stellar core. Although there could be pairing in the strangelets, we believe it would not affect the thermal evolution as they have only a passive role - with the electrons dominating the thermal conduction (the strangelets being analogous to the role of ordinary Ions in traditional crust models).
\subsection{Comparison with Observed Data}
\begin{figure}[!ht]
 \centering
 \vspace{1.0cm}
 \includegraphics[angle=0,width=9 cm]{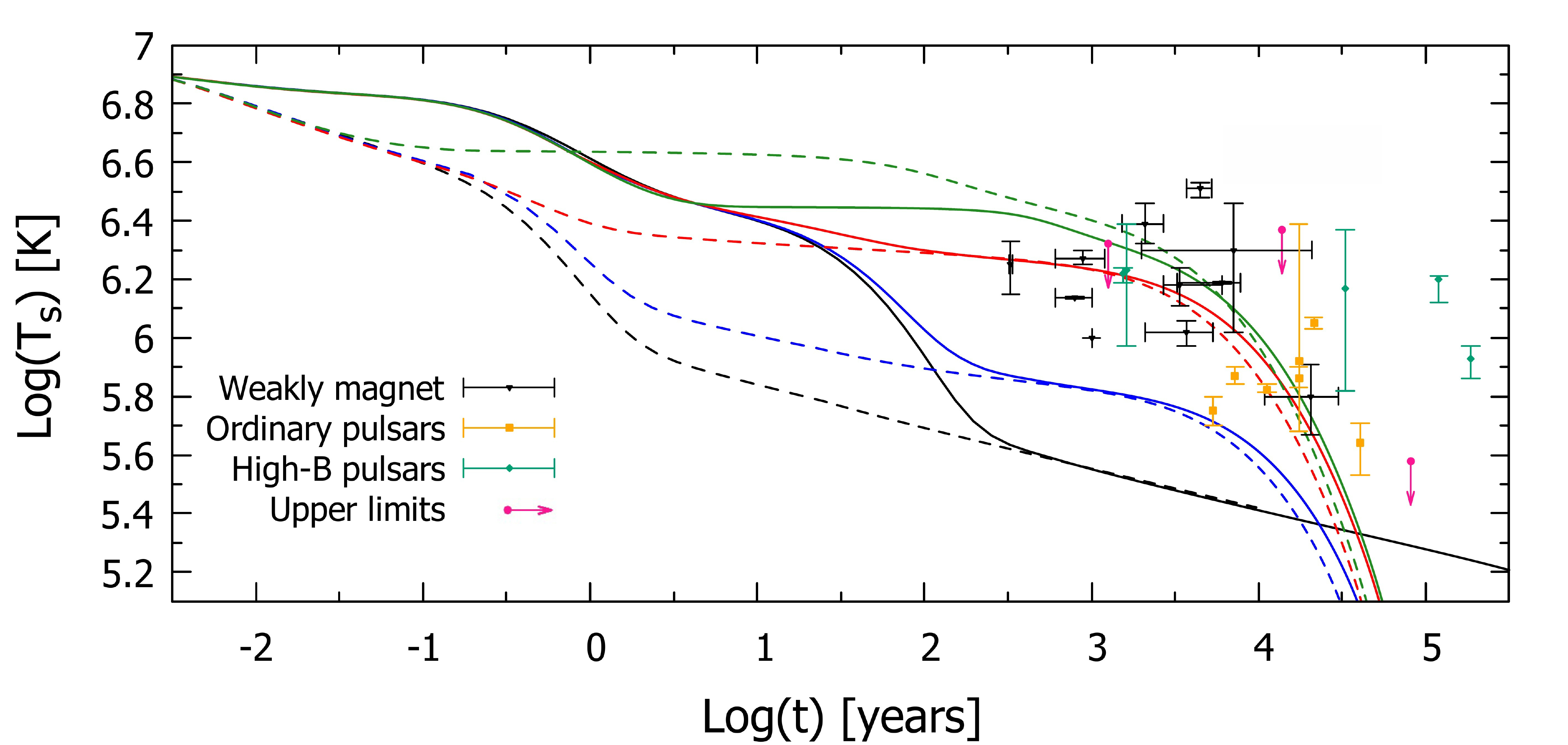}
 \vspace{-0.6cm}
 \caption{\label{fig:data} Same as Fig.~\ref{fig:14}. The curves are
compared with the data from~\cite{potekhin2020thermal}. The data are plotted as indicated in the legend for different neutron star classes. The error bars show uncertainties.}
 \end{figure}

At this moment, we can use our previous results to compare with the current observations. In Figure~\ref{fig:data}, we compare our theoretical results with a set of observed data as described in  Ref.~\cite{potekhin2020thermal}, in which the thermally observable neutron stars are grouped in  different  classes: (i) The Weakly magnetized thermal emitters, that include central compact objects and other thermally emitting isolated neutron stars -- these  mostly emit soft X-ray thermal-like radiation and do not seem to be very strongly magnetized (surface fields below $5\times 10^{11}~G$ or non-determined); (ii)  ordinary pulsars, which comprise thermal data associated with rotation powered pulsars with moderate magnetic fields ($B\sim 10^{12}-10^{13}~G$); (iii) High-B pulsars, objects with strong estimated magnetic fields ($B\sim 10^{13}-10^{14}~G$); and finally; (iv) neutron stars whose temperatures can only be estimated as an upper limit, thought to be associated with relatively young objects (See ~\cite{potekhin2020thermal} for more details). 
 
In Fig.~\ref{fig:data} we show the cooling of $1.4M_{\odot}$ quark stars -- with different pairing gaps -- against the observed data described just above. It quickly becomes evident that without pairing the quark stars cool down too quickly, thus disagreeing with the observed data. Such behavior is not unexpected and has been pointed out in previous works~\cite{Rodrigo2012, Alford2009}. This situation is changed when pairing is included, as the cooling slows down and matches a few of the observed stars. Our results seem to indicate that a moderate pairing with $\Delta \sim 1 - 10$ MeV is favored if the cooling tracks are to go through the data points. At this point, it is opportune to make a few remarks: (i) Fig.~\ref{fig:data} shows that a large set of the data points (mostly in the ordinary pulsars group) lie to the right of the cooling tracks, indicating old objects. One must note however, that unless associated with a supernova remnant (not usually the case for ordinary pulsars, with a few exceptions) one can only estimate the NS age by their spin-down properties. Such estimates should be regarded mostly as an upper limit, as the spin-down age is known to be a very crude estimate (in the few cases in which both spin-down and kinematic ages can be estimated simultaneously they vary drastically); (ii) unfortunately the observed data does not help in differentiating between the nuclear and strangelet crusts studied. As explained in the previous section the difference in the crust composition is more strongly manifested in the process of thermalization of the star, thus, only observation of young stars undergoing such processes (which is not the case with the observed data available) would aid us in differentiating between these models.

\section{DISCUSSION \& CONCLUSIONS}\label{Conclusions}
In this article, we have studied the structure and cooling behavior  of quark stars with two different crust models: (i) nuclear (BPS) matter and, (ii) strangelet crusts. Our goal was to identify possible differences in the cooling behavior of each model as well as to quantify the thermal relaxation properties of quark stars. Quark stars with nuclear crusts were modeled in the traditional manner, assuming an BPS EoS for the crust beginning at the neutron drip density. As for the strangelet crusts we followed the foundations laid in reference ~\cite{Alford2008,Alford2006,Jaikumar2006}, i.e., we consider the possibility that the surface tension of quark matter is low enough to allow for the formation of strangelets. Under this hypothesis, it would be energetically favorable for the quark matter at the low densities of a quark star to rearrange itself into a lattice -- akin to the manner in which the nuclei organize themselves in the traditional crust model for neutron stars. 
As shown in~\cite{Jaikumar2006} strangelet crusts tend to be smaller than their nuclear matter counterpart  with spatial extent  $\sim 20$ m, while a nuclear matter crust has a thickness $\sim 0.5$ km.
Furthermore, according to~\cite{Jaikumar2006}, the small mean free path for electrons scattering off nuggets implies that the thermal conductivity in the crust is much smaller than in the core and
they pointed out that the thermal conductivity of strangelet crusts to be similar to that of nuclear crusts~\cite{Jaikumar2006}. This will influence thermal evolution since the crust will act as an insulator effectively keeping the surface temperature low~\cite{Jaikumar2006, Gnedin2001}. Given such differences we sought to quantify how they manifest themselves in a thermal evolution context.

Our results indicate that most of the thermal differences between the two models studied are manifested in the initial years of cooling. We have found that quark stars with nuclear (and thus thicker) crusts display a slower cooling behavior when compared with QS with strangelets (thinner crusts). Our assessment is that such behavior is mostly due to the difference in crust thicknesses, as the crust acts mostly as a blanket for the initial years of thermal evolution~\cite{Lattimer1994,Gnedin2001,Sales2020}. We have also found that the fact that the crust of the QS's studied is populated with strangelets (as opposed to the traditional ions), does not seem to affect the cooling in any major manner. The reason is that as is the case for the ions in regular NS, the strangelets are mostly inert in the context of thermal processes, with the free electrons being the major agents of heat conduction. In order to quantify our findings  we investigated the thermal relaxation time of quark stars under both models studied. Following the study of~\cite{Sales2020} we have found that the star's relaxation time is linearly dependent on the gravitational mass - with a more sloped curve for the QS with strangelet crust (thus indicating a faster relaxation time). Overall we have found that QS with strangelet crust thermalize in $\sim 1$ year whereas QS with ordinary crust do it in $\sim 100$ years. We have found that this is mostly due to the fact that strangelet crusts are significantly thinner than ordinary hadronic ones. The different mass of the strangelets (in comparisson of ordinary nuclei that compose the crust, also affect the specific heat in the region, although this does not seem to affect the thermal evolution in any major way.

With this work we aimed to investigate the thermal relaxation of quark stars as well as to explore the thermal properties of previously proposed strangelet crust model. We have found that there is a significant decrease in the relaxation time of QS with strangelet crusts (corresponding to a faster thermal evolution). We have also presented the thermal relaxation time of quark stars as a function of their mass, which as far as we know have not been studied before. We currently are expanding this study to consider the effects of rotation and high magnetic field in the structure of the stars we discussed in this work.

\section*{Acknowledgements}
J.Z. acknowledges financial support from CAPES. R.N.
acknowledges financial support from CAPES, CNPq, and
FAPERJ. T.S. acknowledges financial support from CAPES and CNPq. This work is part of the project INCT-FNA Proc. No.
464898/2014-5 as well as FAPERJ JCNE Proc. No. E-26/
203.299/2017. P.J. is supported by the U.S. National Science Foundation Grant No. PHY-1913693.
%

\end{document}